\documentclass[12pt]{article}
\usepackage{natbib}
\usepackage{graphicx}
\usepackage{epsfig}
\usepackage{amssymb}

\begin{document}

\baselineskip24pt

\centerline{\bf Yarkovsky-Driven Spreading of the Eureka Family of Mars Trojans} 

\bigskip
\centerline{Matija \' Cuk$^1$, Apostolos A. Christou$^2$, Douglas P. Hamilton$^3$}

\bigskip

\centerline{$^1$Carl Sagan Center, SETI Institute}
\centerline{189 North Bernardo Avenue, Mountain View, CA 94043} 

\centerline{$^2$Armagh Observatory}
\centerline{College Hill, Armagh BT61 9DG, NI, UK}

\centerline{$^3$Department of Astronomy, University of Maryland}
\centerline{1113 Physical Sciences Complex, Bldg. 415, College Park, MD 20742}

\bigskip

\centerline{E-mail: cuk@seti.org}

\vspace{24pt}
\centerline{Submitted to Icarus}
\centerline{December 3$^{\rm rd}$ 2014.}

\vspace{24pt}

\centerline{Manuscript Pages: 36}

\centerline{Figures: 8}

\centerline{Tables: 1}
\newpage

Proposed Running Head: Eureka Family of Mars Trojans

\vspace{48pt}

Editorial Correspondence to:

Matija \' Cuk

Carl Sagan Center

SETI Institute

189 N Bernardo Ave

Mountain View, CA 94043

Phone: 650-810-0210

Fax: 650-961-7099

E-mail: mcuk@seti.org

\newpage

\noindent {ABSTRACT: Out of nine known stable Mars Trojans, seven appear to be members of an orbital grouping including the largest Trojan, Eureka. In order to test if this could be a genetic family, we simulated the long term evolution of a tight orbital cluster centered on Eureka. We explored two cases: cluster dispersal through planetary gravity alone over 1 Gyr, and a 1 Gyr evolution due to both gravity and the Yarkovsky effect. We find that the dispersal of the cluster in eccentricity is primarily due to dynamical chaos, while the inclinations and libration amplitudes are primarily changed by the Yarkovsky effect. Current distribution of the cluster members orbits are indicative of an initially tight orbital grouping that was affected by a negative acceleration (i.e. one against the orbital motion) consistent with the thermal Yarkovsky effect. We conclude that the cluster is a genetic family formed either in a collision or through multiple rotational fissions. The cluster's age is on the order of 1 Gyr, and its long-term orbital evolution is likely dominated by the seasonal, rather than diurnal, Yarkovsky effect. If confirmed, Gyr-scale dominance of the seasonal Yarkovsky effect may indicate suppression of the diurnal Yarkovsky drift by the related YORP effect. Further study of Mars Trojans is essential for understanding the long-term orbital and rotational dynamics of small bodies in the absence of frequent collisions.}

Key words: asteroids; asteroids, dynamics; planetary dynamics; celestial mechanics.

\newpage 

\section{Introduction}

A Trojan (or "tadpole") coorbital companion is a small body that has the same mean orbital distance as a planet, and librates around the so-called triangular Lagrangian points, which are located 60$^{\circ}$ ahead and behind the planet. Trojans' orbits can in principle be stable for star-planet (or planet-satellite) mass ratios above about 25 \citep{md99}. In our Solar System, only Jupiter, Neptune and Mars are known to have long-term stable Trojan companions \citep{dot08}. Additionally, Saturn's moons Tethys and Dione also have two Trojan coorbitals each \citep{md99, mur05}. Giant planets are thought to have acquired their Trojans during a violent early episode of planetary migration and/or scattering \citep{mor05, nes09, nes13}. Any primordial Saturn and Uranus Trojans would have been subsequently lost through planetary perturbations \citep{nes02, dvo10, hou14}, with the known Uranus Trojans thought to be temporarily captured from among the Centaurs \citep{ale13, fue14}. Some hypothetical Trojans of Earth would have been long-term stable, with the situation at Venus being less clear \citep{tab00, sch05b, cuk12, mar13}; however, so far only temporary coorbitals of these planets are known \citep{chr00, chr11, con11}.

The first Mars Trojan to be discovered was 5261 Eureka in 1990 \citep{bow90}. Since then, a total of nine Mars Trojans have been discovered and were found to be stable \citep[][ and references therein]{mik94, mik94b, con05, sch05, fue13}. The three largest Mars Trojans do not form any kind of cluster: Eureka and 1998~VF$_{31}$ are both in $L_5$ but have very different orbits, and 1999~UJ$_7$ is in $L_4$. Recently, \citet{chr13} proposed that some of the smaller $L_5$ Trojans form an orbital cluster together with Eureka. Subsequently, multiple teams of researchers recognized a likely 6-member orbital cluster \citep[][ A. Christou, 2014, pers. comm.]{fue13}: Eureka, 2001~DH$_{47}$, 2007~NS$_2$, 2011~SC$_{191}$, 2011~SL$_{25}$ and 2011~UN$_{63}$, to which we add 2011~UB$_{256}$ (based on latest orbital elements listed on JPL Solar System Dynamics web-page)\footnote{The recovery of this and other potential Mars Trojans was the result of a targeted campaign by A.~A.~Christou, O.~Vaduvescu and the EURONEAR collaboration \citep{vad13, chr14}.}. In this paper, we will consider these seven objects only, as orbits of more newly discovered objects are likely to have large errors. This is especially true of Mars Trojans' libration amplitudes, which can vary a lot due to relatively small changes in the solutions for their semimajor axes. 

Separately from orbital clustering, spectroscopy can resolve relationships between potential family members. Eureka and 1998~VF$_{31}$ both have high albedos \citep{tri07}, but are not of the same surface composition \citep{riv03}. \citet{riv07} find Eureka to be closest to angrite meteorites, while \citet{lim11} find it to be better matched by olivine-rich R-chondrites. Non-cluster member 1998~VF$_{31}$ is likely to be a primitive achondrite \citep{riv07}, while the sole $L_4$ Trojan 1999~UJ$_7$ has a much lower albedo, and presumably very different composition \citep{mai12}.

In this paper, we explore if the Eureka cluster's orbital distribution could result from a initially compact collisional family spreading due to planetary perturbations and the radiative Yarkovsky effect.

\bigskip

\section{Gravitational Dynamics of Mars Trojans}

\bigskip

Orbits of asteroid families born in collisional disruptions spread due to both gravitational and non-gravitational (usually radiative) forces \citep{bot01}. In the main belt, the Yarkovsky effect \citep{far98, far99, bot06} is by far the most important radiative force on the observable asteroids. The details of a collisional family dispersal are likely to be different among Mars Trojans from those of main-belt asteroids (MBAs). The coorbital relationship with Mars prevents the Trojans from drifting in semimajor axis, making their libration amplitudes, eccentricities and inclinations the only relevant parameters in which we can identify potential families. Additionally, they are largely free of collisions, allowing for the YORP radiation torques \citep{rub00, bot06} to fully dominate the evolution of their spins, with implications for the long-term behavior of the Yarkovsky drift (sub-km Mars Trojans are expected to have their spins completely re-oriented by YORP in a less than a Myr). All these factors make it hard to use the lessons from MBA families for studying a potential family among Mars Trojans, and independent numerical modeling is clearly needed. In order to separate the effects of (purely gravitational) planetary perturbations and the radiation forces, we decided to first model spreading of a family due to gravity alone. Such a simulation is certainly unlikely to reflect a real-life Mars Trojan family consisting mostly of sub-km bodies, but is valuable in providing a control for our Yarkovsky simulations. 

We used the SWIFT-rmvs4 symplectic integrator which efficiently integrates perturbed Keplerian orbits, and is able to resolve close encounters between massless test particles and the planets \citep{lev94}. While in previous versions of SWIFT the timestep used for integrating planets depended on the timing of particle-planet encounters, SWIFT-rmvs4 propagates planets using constant timestep, not dependent on the fate of test particles. This enabled all of the 100 test particles to experience the same history of the chaotic inner Solar System, despite the computation being divided between five different processors. The initial conditions for the eight planets and Eureka are based on the vectors for January 1st, 2000, downloaded from JPL's horizons ephemeris service.\footnote{Vectors used for test-particle simulations were obtained in 2013, while those used to produce Table \ref{cluster} are from August 2014, leading to slight differences between the orbit of Eureka and the centers of the simulated clusters.} Test particles had the same initial positions as Eureka, with their velocities differing slightly from that of Eureka. Small kicks to y and z components of the particle's velocity (in the ecliptic coordinate system) were assigned according to a 5x20 grid. The size of the grid step was $10^{-4}$ of the relevant velocity component, amounting to 1 m/s in the y-direction, 0.5 m/s in the z-direction (comparable to Eureka's likely escape velocity). This was in excess of a realistic collisional fragment dispersion, but enabled us to sample a larger phase space (for comparison, escape from the Trojan region would require about 30 m/s). The simulations were run for $10^9$ years with a 5-day timestep. 

At the end of the simulation, 98 of the hundred particles were still Mars Trojans (the remaining two were destabilized). This agrees with the results of \citet{sch05}, who find that Eureka is most likely long-term stable, with only $\simeq 20\%$ chance of escape over 4.5 Gyr. Also, just like \citet{sch05}, we find that it is the eccentricity that disperses most rapidly due to planetary perturbations (Fig. \ref{newfig1}). The eccentricity dispersion of our synthetic cluster reaches that of the actual cluster in at most a few hundred Myr. This is contrasted by the much slower dispersion in inclination; the inclination dispersion of the synthetic family does not approach the size of the Eureka cluster by the end of the integration. This discrepancy between the eccentricity and inclination dispersals is independent of which bodies we include in the Eureka cluster. If we exclude outlying 2001~SC$_{191}$ from the cluster, inclinations could be explained by a gravity-only dispersal over the Solar System's age. However, the age of the family according to the scatter of member eccentricities would be only $10^8$ years or so, clearly indicating that this cannot be a collisional cluster that has been spreading due to gravity alone (Fig. \ref{newfig2}).

Our choice of initial conditions produced a relatively large dispersal in libration amplitudes ($2-20^{\circ}$), which has not changed appreciably in the course of the simulation. As we noted, our grid was too large for a realistic collisional family, as some test particles received initial kicks as large as 6~m/s, much in excess of Eureka's escape velocity. Overall, it appears that long-term planetary perturbations do not affect libration amplitudes as much as they do eccentricities, as found by \citet{sch05}. This relatively rapid dispersal of eccentricities is the most important lesson from the simulations described in this section, and is certainly relevant for analysis of more realistic simulations including the Yarkovsky effect.

\bigskip

\section{Integrations including the Yarkovsky drift}

\bigskip

When introducing radiation forces into our simulation, we decided to restrict ourselves to the simplest case: a constant tangential force on the particle. This is a good model of the net Yarkovsky acceleration over short timescales, but may not be accurate in the long term if the spin axis or the spin rate are changing. In the absence of collisions, spin evolution of a Trojan is expected to be dominated by the YORP torque \citep{rub00, bot06}. Depending on the model, YORP torques may evolve bodies toward asymptotic states \citep{vok02, cap04}, or could cause reshaping that would settle the body into a quasi-stable state \citep{sta09, cot13}. Depending on the outcome of YORP, a sub-km asteroid not subject to collisions could keep the same sense of rotation indefinitely, or would cycle through obliquities and spin rates on 1-Myr timecales. This has clear implications for Yarkovsky effect over Gyr timescales. Therefore our constant Yarkovsky force should be seen as representing a long-term average of the thermal drift. Regardless of the actual behavior of the body, the average thermal drift rate over 1 Gyr is a single number, and by giving our particles a range of Yarkovsky drift rates we are effectively exploring this averaged rate. 

Ninety-six clones of Eureka were assigned drift rates in the range $-1.4 \times 10^{-3}$ AU/Myr $< \dot{a} < 1.4 \times 10^{-3}$~AU/Myr and integrated for one billion years. While Yarkovsky-capable standard integrators are available \citep{bro06}, we wanted to have full control over our numerical experiment and used the home-made SIMPL code \citep{cuk13}. SIMPL does not allow for close encounters, but stable Mars Trojans never have encounters with Mars or other planets. We used the parametric migration option in SIMPL to assign a constant tangential force on the particle (i.e., a force in the plane of the orbit, perpendicular to the heliocentric radius vector). The force was assigned $r^{-3.5}$ dependence on heliocentric distance, in order to produce $\dot{a} \sim a^{-2}$ typical of the Yarkovsky effect \citep{bot06}. We also modified SIMPL so that our Mars Trojans are treated as massless particles. The timestep was 9.1 days.

Results of this first Yarkovsky simulation are shown in Figs. \ref{newfig3} and \ref{newfig4}. In all figures, particles with inward-type Yarkovsky drift (which can be interpreted as retrograde rotators) are represented with red plusses, and those with outward-type drift (which have to be prograde rotators) with blue Xs. Note that the direction of migration discussed here is one bodies would follow if they were not co-orbitals; in our simulation the mean semimajor axes of test particles always stays equal to that of Mars as must be the case for coorbital motion (unless an escape occurs). Also, the connection between Yarkovsky drift direction and rotation holds only for diurnal Yarkovsky effect, while the seasonal variety makes all bodies migrate inward; this will be discussed in more detail in Section 4. The top panel in Fig. \ref{newfig3} shows evolution of the particles' eccentricities over time. There is a systematic trend that outward migrators increase their eccentricities, while the inward ones decrease their eccentricities over time. These two groups are not well-separated and there is lots of overlap, likely due to gravitationally-generated chaos. 

\begin{figure*}[h]
\includegraphics{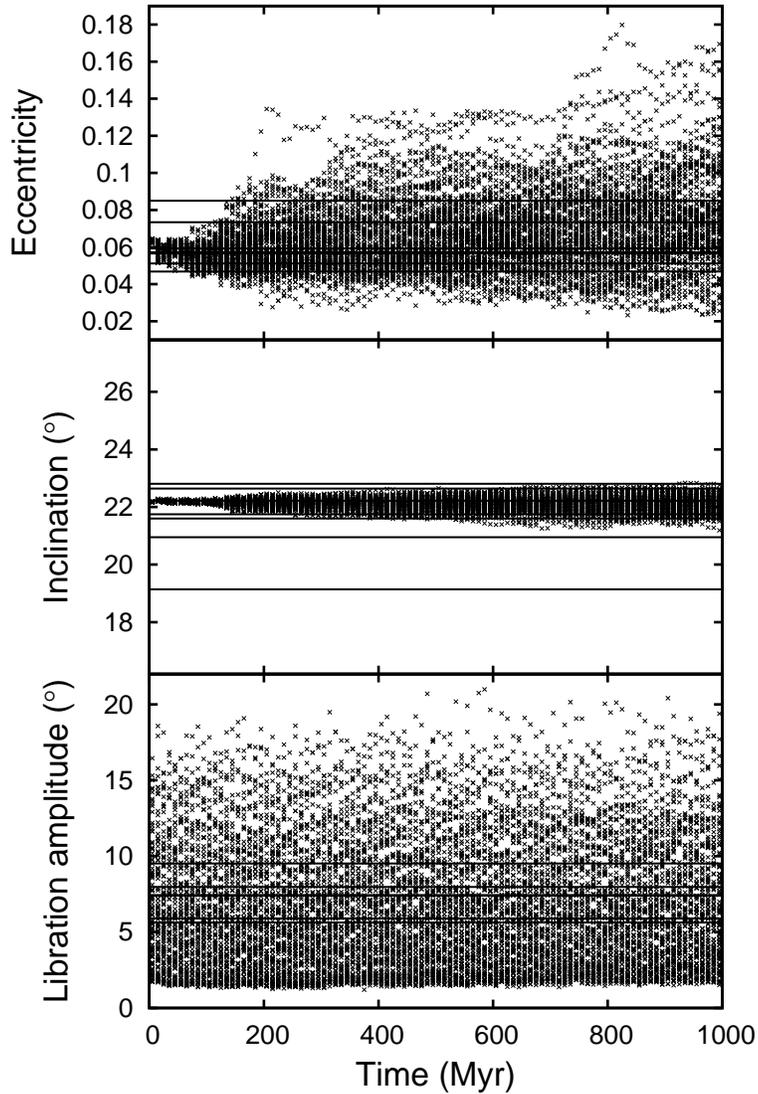}  
\caption{Evolution of the 100 Mars Trojan test particles' eccentricities (top), inclinations (middle) and libration amplitudes (bottom) in the 1 Gyr simulation which included only gravitational forces. Each point plots a mean value calculated over 10 Myr for each particle. Horizontal bars show mean elements for known cluster members.} 
\label{newfig1}
\end{figure*}

\begin{figure*}[h]
\includegraphics[]{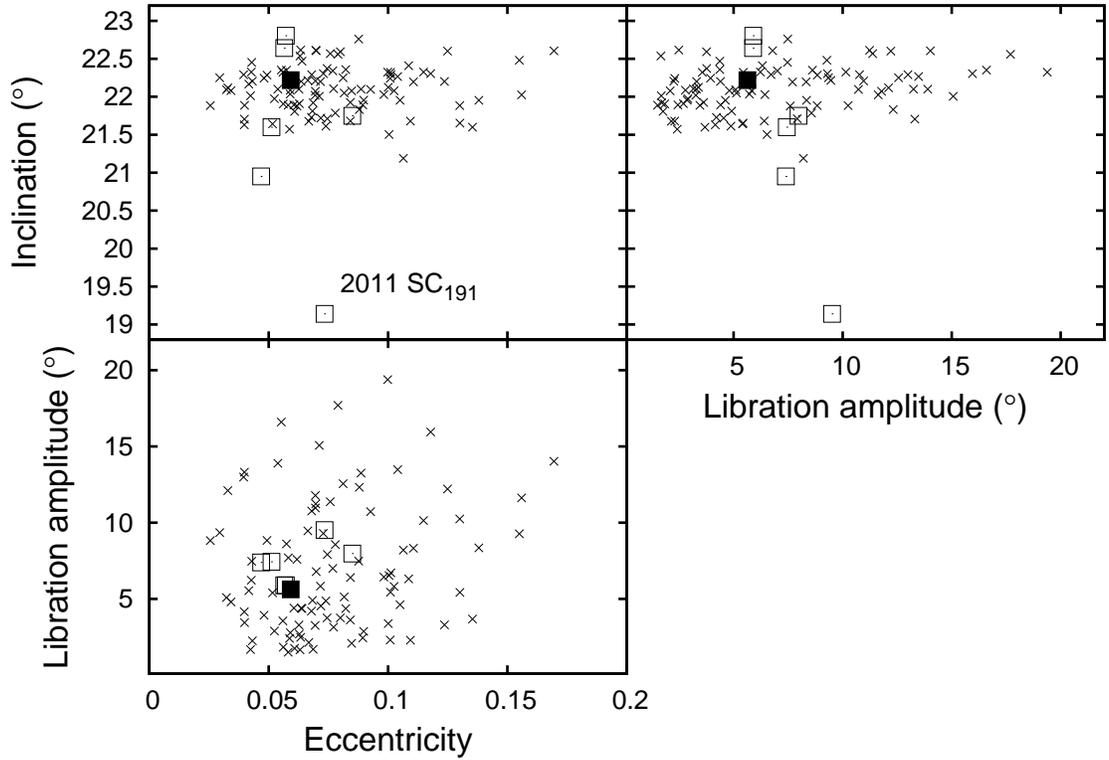}  
\caption{Correlations between the test particles' eccentricities and inclinations (top left), eccentricities and libration amplitudes (bottom) and inclinations and libration amplitudes (top right) at the end of the gravity-only 1 Gyr simulation. All of the elements were averaged over the last 10 Myr. Known family members are plotted as open squares, with Eureka shown as a filled square.} 
\label{newfig2}
\end{figure*}

\begin{figure*}[h]
\includegraphics[]{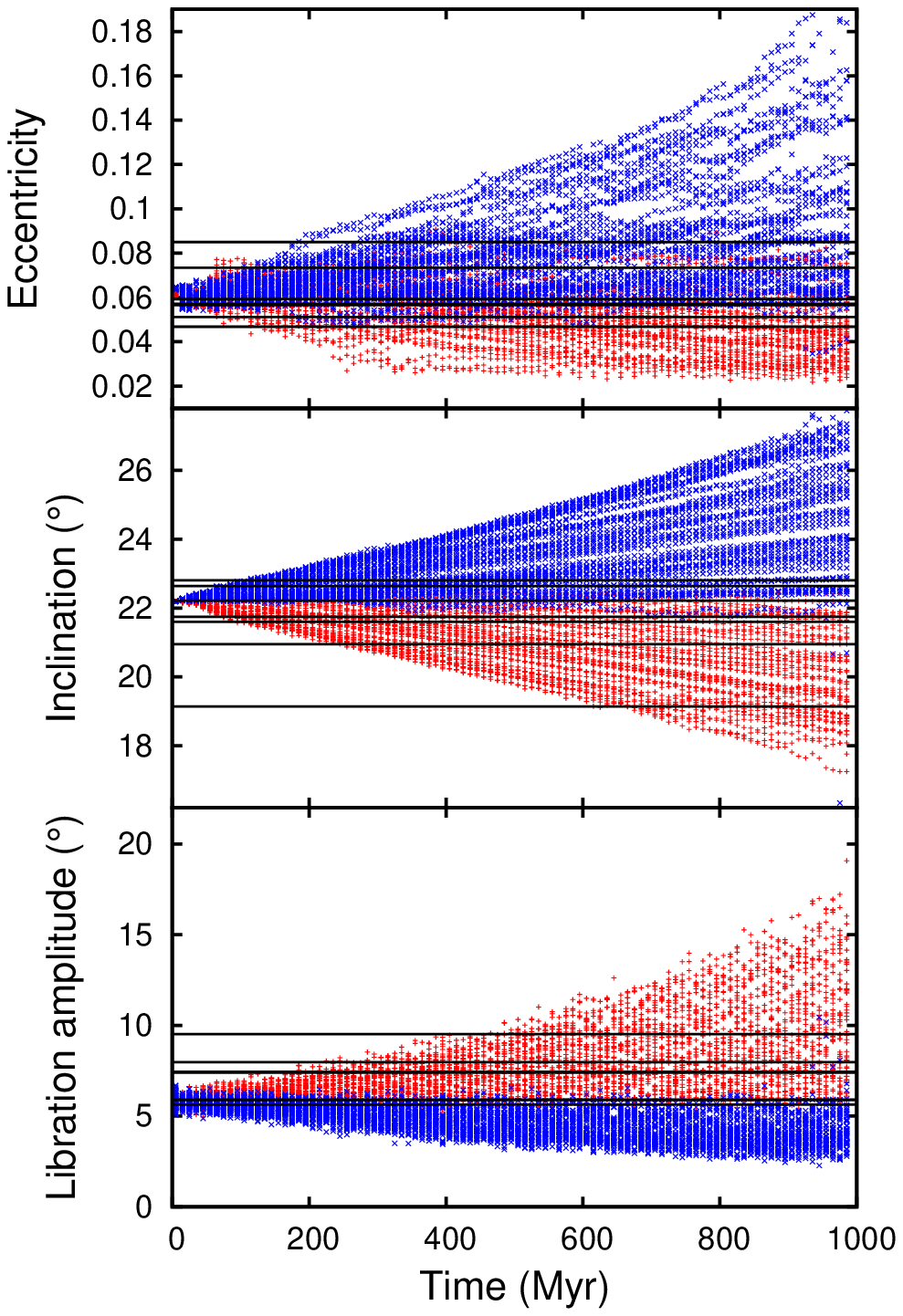}  
\caption{Evolution of the 96 Mars Trojan test particles' eccentricities (top), inclinations (middle) and libration amplitudes (bottom) in the 1 Gyr simulation which included the Yarkovsky effect. Each point plots a mean value calculated over 10 Myr for each particle. Test particles with inward-type Yarkovsky drift are plotted as red pluses, those with outward-type drift are plotted as blue x's. Horizontal bars show mean elements for known cluster members.} 
\label{newfig3}
\end{figure*}

\begin{figure*}[h]
\includegraphics[]{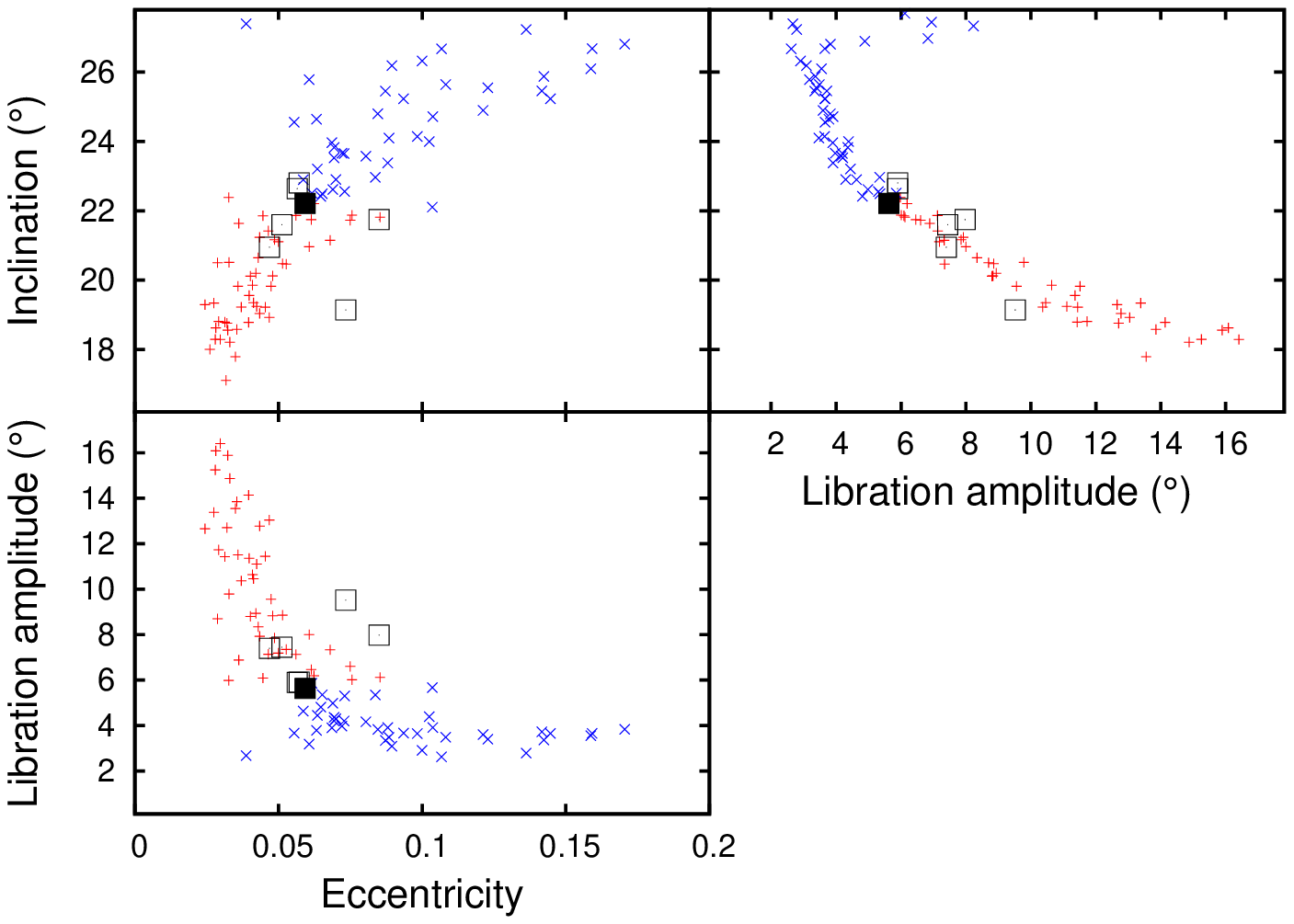}  
\caption{Correlations between eccentricities and inclinations (top left), eccentricities and libration amplitudes (bottom) and inclinations and libration amplitudes (top right) at the end of the 1 Gyr simulations incorporating the Yarkovsky effect. The elements have been averaged over the last 10 Myr. Test particles with inward-type Yarkovsky drift are plotted as red pluses, those with outward-type drift are plotted as blue x's. Known family members are plotted as open squares, with Eureka shown as a filled square.} 
\label{newfig4}
\end{figure*}

Comparison with Fig. \ref{newfig1} shows that dispersion of eccentricities, even for outward migrators, is somewhat suppressed in the Yarkovsky simulation. This is easy to understand for would-be inward migrators, as their eccentricities are directly decreased by the Yarkovsky effect. However, even the particles with positive or near-zero Yarkovsky drift have more stable eccentricities in the second simulation. The most likely explanation is that, due to the well-known chaotic nature of planetary orbits, the amount of chaotic diffusion was smaller in the Yarkovsky simulation as those particles experienced a different dynamical history. The top panel in Fig. \ref{newfig5} shows the eccentricity of Mars in the two numerical experiments, and it is clearly lower for much of the Yarkovsky simulation. 

In order to get a better grip on the effect of martian eccentricity, we ran seven more planets-only 1000 Gyr simulations (i.e. without Trojans). Initial conditions were very similar to our Yarkovsky simulation (shown in Fig. \ref{newfig3}), only with Earth initially shifted by few hundred meters along the x-axis. The bottom panel of Fig. \ref{newfig5} shows the (averaged) eccentricity of Mars in these simulations (together with that from the original Yarkovsky run). It is clear that the test particles plotted in Fig. \ref{newfig3} experienced below-average eccentricity of Mars' orbit over 1 Gyr. We then chose one of the higher-eccentricity simulations (plotted with thick dashed line), and re-ran it with the same Trojan test-particles as in the first Yarkovsky simulation. The results of this second Yarkovsky simulation are plotted in Figs. \ref{newfig6} and \ref{newfig7}. The Trojans' final eccentricities in Fig. \ref{newfig7} are more spread out than in Fig. \ref{newfig4}, confirming that the eccentricity dispersion of the family depends on the history of martian eccentricity. However, both simulations are equally valid, and we cannot say (except probabilistically) what the orbit of Mars was doing over 100 Myr or longer timescales \citep{las11}. Therefore the eccentricity is not very useful for determining the history of age of the family, and we must rely on inclinations and libration amplitudes which are less chaotic.

The middle panels in Figs. \ref{newfig3} and \ref{newfig6} show the evolution of inclinations over time. The inclinations vary much more than in gravity-only simulations, with all of the inward-type migrators decreasing their inclinations, while all of the outward migrators have growing inclinations. This is consistent with results of \citet{lio95} for Poynting-Robertson drag, and happens because the Yarkovsky effect removes (or adds) angular momentum in the plane of the orbit, while the restoring force comes from Mars, which is less inclined than the Eureka cluster Trojans. The drift in inclination appears much more orderly and linear than that in eccentricity, as we already know that inclination is quite stable in the absence of the Yarkovsky effect (Fig. \ref{newfig1}). The range of inclinations at the end of 1 Gyr matches the spread of inclinations within the Eureka cluster, with the peculiarity that all of the cluster members appear to have been migrating in one direction. The bottom panels show the evolution of the test-particles' average libration amplitudes. They also appear to be affected primarily by the Yarkovsky drift, with outward-type migrators having shrinking libration amplitudes, while the inward-type migrators have their libration amplitudes grow. Just as for inclination, cluster members (other than Eureka) show signs of a force that would have caused inward migration.

Configurations at the end of our Yarkovsky simulations are shown in Figs \ref{newfig4} and \ref{newfig7}. The distribution of the true cluster eccentricities does not follow the synthetic cluster in Fig. \ref{newfig4}, with the two high-eccentricity bodies being outliers. The second Yarkovsky simulation fares somewhat better (Fig. \ref{newfig7}), but 2011~SC$_{\rm 191}$ is still an outlier in eccentricity (as well as being most distant from Eureka in inclination). On the other hand, both synthetic clusters match the real one in the inclination-libration amplitude plots (top right panels of Figs. \ref{newfig4} and \ref{newfig7}). Those cluster members that are well-separated from Eureka in their orbital elements (especially 2011~SC$_{\rm 191}$) appear to have been trying to migrate inward over the age of the cluster. The simplest conclusion from these results is that Eureka cluster is indeed a genetic family, and that at least some of the cluster members show the effects of sustained {\it inward} Yarkovsky drift over Gyr timescales. 

\bigskip

\section{Discussion}

\bigskip

Before discussing the age of the family, we should consider the issue of the family originating in a single event, rather than having fragments ejected from Eureka (or each other) at different times. When all three orbital parameters are very similar (as in the case of 2001~DH$_{\rm 47}$ and 2011~UB$_{\rm 256}$), a very recent breakup (either rotational or collisional) is usually indicated \citep{vok08}. If some of the family members are more recent fragments, we cannot make any statements about their Yarkovsky behavior (or more accurately, lack thereof), as they may not have had the time to evolve. So the "age of the family" that we discuss here is the maximum age of the family, based on the dispersal of the most distant members (which are likely to have been the first to separate). 

From comparison with test particles, 2011~SC$_{\rm 191}$ should have experienced a Yarkovsky force equivalent to inward migration at 7.7$\times 10^{-4}$~AU/Myr, if it were to migrate from Eureka in 1 Gyr. If we assume it has albedo of 0.4 \citep{tri07}, this would made its diameter $D=300$~m. We can compare that with rates of Yarkovsky drift computed by \citet{bot06}. Since $\dot{a} \sim a^{-2}$, rate of 7.7$\times 10^{-4}$~AU/Myr at 1.52 AU is equivalent to the rate 2.8 $\times 10^{-4}$~AU/Myr at 2.5 AU \citep[the benchmark distance used in][ Figure 5]{bot06}. Comparing this estimate to the theoretical curves in the same figure, we find that the inferred drift rate for 2011~SC$_{191}$ is firmly within the range of predictions for a 300~m body. The same calculation for somewhat larger (and less divergent) 2007~NS$_2$ gives a similar result. Somewhat smaller 2011~SL$_{25}$ and 2011~UN$_{63}$, which have evolved about the same amount in inclination and libration amplitude as 2007~NS$_2$, may imply a slower-than-predicted rate, but uncertainties in predictions themselves are to big to make any firm conclusions. 

All of the above comparisons assumed the age of the family to be 1 Gyr. The fact that that assumption produces reasonable values for the Yarkovsky drift may imply that the real age is comparable. However, estimates of Yarkovsky drift vary by a factor of a few depending on asteroid properties \citep{bot06}, and we do not know enough about the family members to further refine these estimates. Evolution of the spin axis through YORP could lead to much slower effective Yarkovsky drift, meaning that the family could be significantly older. 

\begin{figure*}[h]
\includegraphics[]{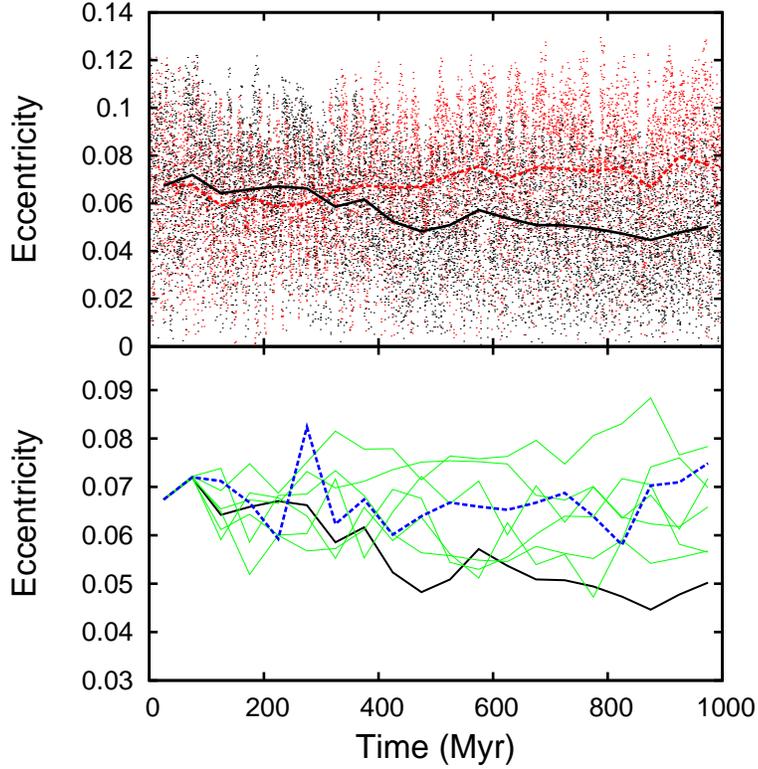}  
\caption{Top: Evolution of the Martian eccentricity over 1 Gyr gravity-only simulation (red dots) and the first 1 Gyr Yarkovsky simulation shown in Figs. \ref{newfig3}-\ref{newfig4} (black dots). Values averaged over 50 Myr bins are shown as red dashed and black solid lines for the non-Yarkovsky and the Yarkovsky simulations, respectively. Bottom: Eccentricity of Mars in eight simulations with similar initial conditions, averaged over 50 Myr bins. The thick solid black line plots the original Yarkovsky simulation (same as solid black line in the top panel), while the remaining lines plot its "clones", including the case used for the Yarkovsky simulation shown in Figs. \ref{newfig6} and \ref{newfig7} (thick blue dashed line).} 
\label{newfig5}
\end{figure*}

\begin{figure*}[h]
\includegraphics[]{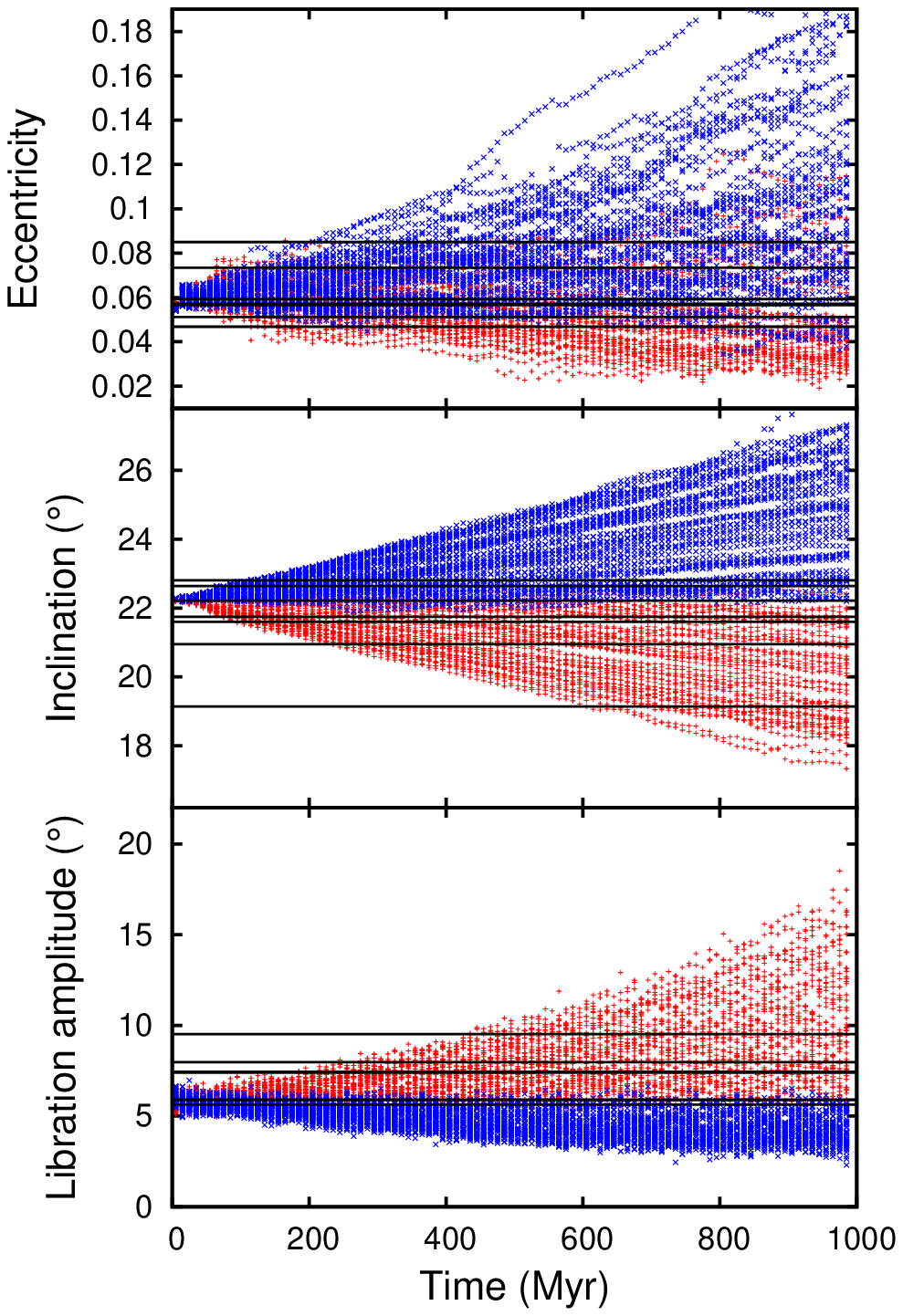}  
\caption{Evolution of the 96 Mars Trojan test particles' eccentricities (top), inclinations (middle) and libration amplitudes (bottom) in the second 1 Gyr simulation including the Yarkovsky effect. Each point plots a mean value calculated over 10 Myr for each particle. Test particles with inward-type Yarkovsky drift are plotted as red pluses, those with outward-type drift are plotted as blue x's. Horizontal bars show mean elements for known cluster members.} 
\label{newfig6}
\end{figure*}

\begin{figure*}[h]
\includegraphics[]{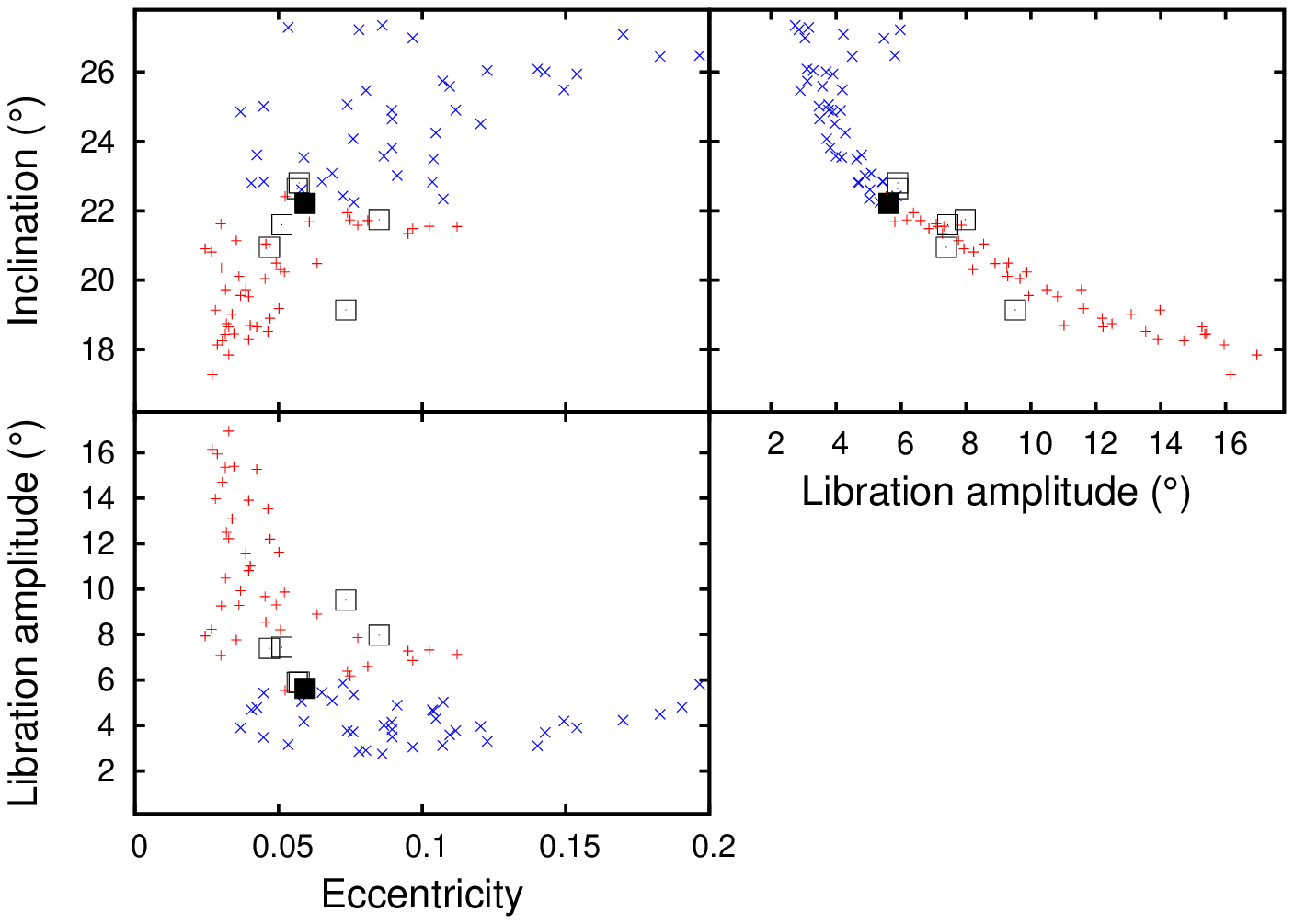}  
\caption{Correlations between eccentricities and inclinations (top left), eccentricities and libration amplitudes (bottom) and inclinations and libration amplitudes (top right) at the end of the second 1 Gyr simulations incorporating the Yarkovsky effect. The elements have been averaged over the last 10 Myr. Test particles with inward-type Yarkovsky drift are plotted as red pluses, those with outward-type drift are plotted as blue x's. Known family members are plotted as open squares, with Eureka shown as the solid square.} 
\label{newfig7}
\end{figure*}

In Fig. \ref{newfig4}, the eccentricities of bodies with zero or inward-type Yarkovsky drifts do not spread enough over 1 Gyr to match the dispersal of the family. Therefore, an age of the family longer than 1 Gyr would be consistent with this particular simulation. On the other hand, the difference between the eccentricity dispersions of test particles in Figs. \ref{newfig2}, \ref{newfig4} and \ref{newfig7} is due to the lower eccentricity of Mars in the first Yarkovsky simulation. Eccentricities of particles in Fig. \ref{newfig3} disperse fast until 300-400 Myr, when the dispersion rate (especially for would-be inward migrators) slows down. This correlates well with the evolution of Martian eccentricity, as we see the eccentricity in the first Yarkovsky simulation (black solid lines inn Fig. \ref{newfig5}) dropping below that in the gravitational and second Yarkovsky simulations (dashed red and blue lines, respectively, in Fig. \ref{newfig5}) at about 300~Myr. Of course, this divergence is not in any way related to the Yarkovsky effect, but to slightly different numerical integrators, timesteps or initial conditions used for the two simulations. While all of these integrations are equally valid in principle, additional planetary simulations indicate that higher martian eccentricity is more likely than that seen in the first Yarkovsky simulation (bottom panel in Fig. \ref{newfig5}). This is in agreement with previous work, for example the large sets of simulations done by \citet{las08}. 

Therefore, the age of the family based on eccentricity spreading is {\it on average} consistent with about 1 Gyr, but with very large uncertainties. Note that 2011~SC$_{\rm 191}$, apart from being most evolved through Yarkovsky drift, is also an outlier in eccentricity, even in the second Yarkovsky simulation (which used a more realistic martian eccentricity). Such a correlation is likely if the family members have different ages, with 2011~SC$_{\rm 191}$ being presumably the oldest extant fragment. This would be an argument for the origin of the family in a series of YORP disruptions, similar to those that produced asteroid pairs \citep{pra10}.  

The fact that the four family members that are well-separated from Eureka in the orbital-element space all show signs of inward-type Yarkovsky migration is not necessarily significant and may be due to chance. If future family members are found to exhibit outward-type average drift, the observed dispersion would be consistent with the diurnal Yarkovsky effect. However, lack of any outward-type migrators could also be the sign of the dominance of the seasonal Yarkovsky effect. Using the theory of \citet{vok99}, we find that the predicted drift rate for a $D=300$~m body at 1.5~AU should be $-6 \times 10^{-4}$~AU/Myr (assuming a $90^{\circ}$ obliquity), very similar to $-7.7 \times 10^{-4}$~AU/Myr we infer for 2011 SC$_{191}$ when assuming the Eureka family age of 1 Gyr. Here we also assumed that the thermal parameter $\Theta=1$, but otherwise this estimate is not affected by thermal properties of the asteroid, as the Yarkovsky seasonal effect relies on the heat from absorbed sunlight penetrating to depths much smaller than the size of the body \citep{vok99}. The presence or absence of regolith cannot therefore be inferred, unless the diurnal Yarkovsky effect can also be observed. Discoveries of more Trojans, and establishing the population outward-type migrators would be the best way of determining relative importance of the two Yarkovsky effects. 

The slopes of the synthetic families in top right panels of Figs. \ref{newfig4} and \ref{newfig7} match that of the real family rather well, indicating the type of radiation force included. We tested functional forms of the acceleration other than $r^{-3.5}$ (Fig. \ref{newfig8}) and they do not lead to the same slope (although a somewhat steeper radial dependence like $r^{-5}$ cannot be excluded at this point). In particular, we find that a constant negative tangential acceleration (leftmost lines in Fig. \ref{newfig8}) produces a {\it decrease} in the libration amplitude \citep[this is equivalent to the results of ][ for constant outward migration of the planet]{fle00}, rather than an increase which is apparent among Eureka family members. However, since both diurnal and seasonal Yarkovsky effects have the same distance dependence \footnote{This radial dependence of the seasonal Yarkovsky is true for asteroids but not for much cooler Trans-Neptunian Objects \citep{bot06}.}, there is no way to use this slope to distinguish between these two variants of the Yarkovsky effect.

The closest match to Mars Trojans in terms of sizes and albedos (if not the collisional environment) would be sub-km Hungaria family asteroids \citep{war09}. \citet{war09} show a distribution of known Hungaria family members, and it appears that at 18th magnitude, where incompleteness sets in, more bodies populate the inner part of the family. Of course, Hungarias are subject to observational bias that strongly favors closer-in bodies. However, if this trend of inward-type migration perseveres as more Eureka family members are known, it would be interesting to revisit this issue for small Hungarias, which offer a much larger, and yet similar, sample.

\bigskip

\section{Summary and Conclusions}

In this paper, we have reached the following conclusions:

1. The Eureka cluster of Mars Trojans is a genetic family, most likely originating on (5261) Eureka. While the family could have formed in a single event, separation of different members at different times due to YORP breakups may be more likely.

\begin{figure*}[h]
\includegraphics[]{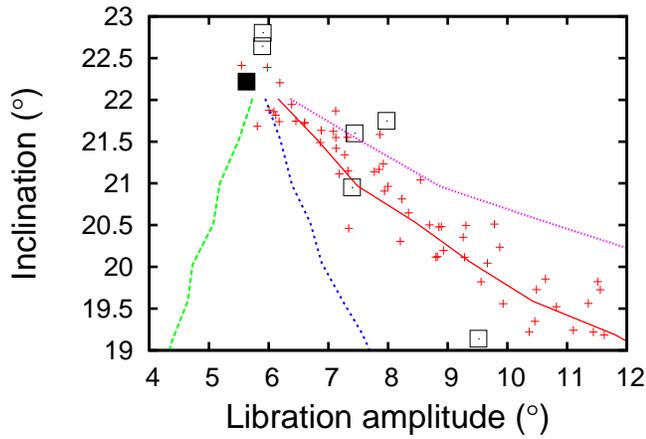}  
\caption{Correlations between inclinations and libration amplitudes during 10 Myr simulations using different orbital migration forces (with a greatly accelerated drift rate of -0.15~AU/Myr). Clockwise from top right, the four particle tracks used negative tangential acceleration with radial dependence of $r^{-5}$, $r^{-3.5}$, $r^{-2}$ and $r^0$ (i.e. constant force), respectively. On these particle tracks, each point is an average over 1 Myr. Test particles with inward-type Yarkovsky accelerations from the two 1-Gyr simulations including the Yarkovsky effect are also plotted as red pluses, and known family members are plotted as open squares, with Eureka shown as the solid square.} 
\label{newfig8}
\end{figure*}

2. Inclinations and libration amplitudes of Eureka family have spread through the Yarkovsky effect. The shape of the family in the orbital parameter space (Fig. \ref{newfig8}) indicates that the force spreading the family has radial dependence close to $\simeq r^{-3.5}$ expected for the Yarkovsky effect.

3. The age of the family is likely on the order of 1 Gyr. This age is consistent with the expected rates of Yarkovsky drift, but we do not know enough about family members to constrain it further. Spreading of eccentricities is driven by planetary chaos, and the rate of this spreading can vary significantly from one simulation to another. We think that the observed spread in eccentricities is consistent with a $\simeq$1 Gyr age. If the family is a result of sequential YORP breakups, the most distant member 2011~SC$_{\rm 191}$ would likely be older than 1 Gyr, while the remaining members would be significantly younger. 

4. The family spreading in inclination and libration amplitude may be dominated by the seasonal, rather than the better-known diurnal, Yarkovsky effect. The seasonal effect always removed orbital energy, and no family members known so far show signs of outward-type migration (indicative of the diurnal Yarkovsky effect acting on prograde rotators). Having more known family members would answer the question of diurnal vs. seasonal Yarkovsky effect, and would help us understand how the Yarkovsky effect acts on sub-km asteroids in general.

\bigskip

\vspace{48pt}

{\centerline{\bf ACKNOWLEDGMENTS}}

\bigskip

M\' C is supported by NASA's Planetary Geology and Geophysics (PGG) program, award number NNX12AO41G. Astronomical research at the Armagh Observatory is funded by the Northern Ireland Department of Culture, Arts and Leisure. M\'C thanks Hal Levison for sharing his SWIFT-rmvs4 integrator.

\newpage

\bibliographystyle{}

\vspace{3cm}

\begin{table}[h]
\begin{center}
\caption{Dynamical properties and absolute magnitudes H of known Eureka cluster members. Eccentricities and inclinations are mean values computed over $10^7$~yr, using initial conditions from JPL Solar System Dynamics site, retrieved on 08/15/2014. Here and elsewhere in the paper, the mean libration amplitudes are computed as ${\pi \over 2n} \sum_1^n | \lambda_M-\lambda-60^{\circ}|$ (summed over output intervals), where $\lambda$ is mean longitude, and subscript M refers to Mars. All inclinations in this paper are measured relative to the J2000 ecliptic. Assuming an albedo of $0.4$, absolute magnitudes of $H=16$ and $H=19$ correspond to diameters of $D=1.3$~km and $D=0.33$~km, respectively.\label{cluster}}
\bigskip
\begin{tabular}{|l|c|c|c|c|}
\hline\hline
Trojan & Amplitude ($^{\circ}$)& Eccentricity & Inclination ($^{\circ}$) & H (mag)\\
\hline\hline
(5261)~Eureka & 5.63 & 0.0593 & 22.22 & 16.1\\     
\hline
2001~DH$_{47}$ & 5.90 & 0.0572 & 22.80 & 18.7\\     
\hline
2007~NS$_2$ & 7.40 & 0.0468 & 20.95 & 18.1\\     
\hline
2011~SC$_{191}$ & 9.52 & 0.0734 & 19.14 & 19.3\\
\hline
2011~SL$_{25}$ & 7.97 & 0.0850 & 21.75 & 19.4\\         
\hline
2011~UB$_{256}$ & 5.89 & 0.0565 & 22.64 & 20.1\\     
\hline
2011~UN$_{63}$ & 7.44 & 0.0512 & 21.60 & 19.7\\
\hline
\end{tabular}
\end{center}
\end{table}

\end{document}